\documentclass{scrartcl}

\usepackage{color,array,amsthm}
\usepackage{graphicx}
\usepackage{amsmath,amssymb}
\usepackage{xcolor}
\usepackage{hyperref}
\usepackage{tikz}
\title{Statistically Robust Resource Block Allocation for Satellite Communications}
\author{
  C. Manapragada \and
  L. Decreusefond \and
  P. Martins 
}
\date{2026}
\begin{document}
\maketitle

\begin{abstract}It is critical to dimension (accurately estimate capacity of) a satellite system prior to deployment, as it is very
expensive to reconfigure launched satellite systems that fail to meet demand or that waste capacity. The fundamental requirement is a
dimensioning rule for resource blocks (RBs) given a satellite footprint and a target overload probability (target Quality-of-Service).
The rule must be robust to the spatial covariance structure of signal attenuation, which is generally unknown both at
the time of pre-deployment dimensioning and afterwards.
Existing approaches address parts of this problem, but there does not yet exist a footprint-level
RB dimensioning rule for the satellite context. We develop such a rule: starting with a Gaussian attenuation field that induces a covariance structure
inspired by classical work on spatial covariance of attenuation, we sample users at random along with their field-based attenuation values, and estimate
aggregate RB demand for a target overload probability. We do this in two complementary ways: a Monte Carlo route that gives a simulation-derived RB budget
for a given target overload probability, and a concentration route that gives a conservative analytic upper bound on the target overload
probability for a given RB budget (such as the one obtained through simulation). Taken together, these complementary approaches give a principled way to
dimension RBs for a satellite footprint under spatially correlated attenuation.

\end{abstract}

\noindent\textbf{Keywords:}
Gaussian fields, non-terrestrial networks, resource block dimensioning, satellite communications, spatial correlation.

\section{Introduction}
\label{sec:introduction}
An operator dimensioning a satellite footprint before deployment needs a number: how many resource blocks (RBs) should be available for
that footprint so that aggregate user demand is served with the desired reliability?

This question is becoming more important as non-terrestrial networks are increasingly treated as part of the 5G and 6G communications
ecosystem~\cite{azari2022ntn_survey,giordani2021space}. Broadly, recent satellite-system studies emphasize that traffic demand,
radio-resource allocation, payload limits, and constellation design have to be considered together before
deployment~\cite{zamacola2025joint}. The present work focuses on one key quantity inside that broader problem: the RB number needed
for a satellite footprint.

For satellite systems, both under-dimensioning and over-dimensioning are costly. Under-dimensioning can lead to unmet demand or expensive
post-deployment workarounds, while over-dimensioning ties scarce payload and spectrum resources to capacity that may not be needed.
Because payload and constellation choices are made before launch and can only be corrected after deployment at very high expense through
limited reconfiguration or additional capacity deployment~\cite{esa_gfp}, the RB requirement should be estimated accurately before
deployment, ideally with strong theoretical guarantees on the overload probability. To compute this RB requirement, a model has to
account for spatial attenuation, random user locations, the link-budget-to-RB mapping, and a target overload probability.

The difficulty is that the required resource block (RB) budget is not determined by the distribution over the one-link signal
attenuation~\cite{iturp2108,iturp618} alone.
When attenuation is spatially correlated, nearby users tend to experience favorable or unfavorable conditions together. As a result, aggregate
RB demand can vary significantly even when the marginal attenuation law is fixed. Thus, a dimensioning rule that ignores spatial covariance can
misestimate the required RB budget. At the pre-deployment stage, the spatial covariance of attenuation is therefore an essential part of the RB
dimensioning question rather than a secondary propagation detail.

We therefore model the spatial attenuation field itself, rather than only the one-link marginal distribution of attenuation. The dimensioning rule takes
as inputs a satellite footprint, a random user field, and an operator-specified overload target. Spatial attenuation is represented by a
correlated Gaussian field over the footprint. For a given set of user locations, the field is evaluated at those locations, the resulting
attenuation values are converted into capped per-user RB demands, and the demands are aggregated over the footprint. The output is the
footprint-level RB budget required to satisfy the target overload probability. Because our model is parametric, the impact of individual parameters can
be studied through simulation and analysis.

We compute this budget in two complementary ways. The first is a Monte Carlo route, which estimates the required RB budget from repeated
sampled user locations and a Gaussian vector of attenuation values at those locations (each sampled Gaussian vector is a representation of a latent Gaussian field).
The second is a concentration route, which computes the mean aggregate RB demand induced by each sampled field representation and then bounds the
probability that demand exceeds a proposed RB budget conditional on that representation. This gives both a simulation-derived RB budget and a conservative
analytic comparator for the same target overload probability. Taken together, these complementary approaches may be used for principled pre-deployment
RB dimensioning for a satellite footprint.

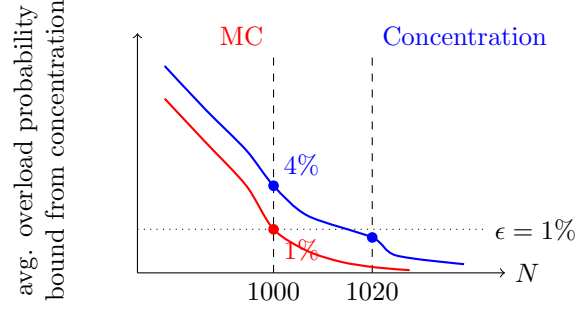
\begin{figure}[t]
\centering
\begin{tikzpicture}[x=0.9cm,y=0.72cm,font=\small]
\draw[->] (0,0) -- (5.4,0) node[right] {$N$};
\draw[->] (0,0) -- (0,4.4);
\node[rotate=90,align=center] at (-1.45,2.25) {avg. overload probability\\bound from concentration};
\draw[dotted] (0,0.8) -- (5.1,0.8) node[right] {$\epsilon=1\%$};
\draw[dashed] (2,0) node[below] {$1000$} -- (2,4.0);
\draw[dashed] (3.45,0) node[below] {$1020$} -- (3.45,4.0);
\node[above left,red] at (2,4.0) {MC};
\node[above right,blue] at (3.45,4.0) {Concentration};
\draw[thick,blue] plot[smooth] coordinates {(0.4,3.8) (1.0,3.0) (1.6,2.25) (2.0,1.6) (2.5,1.05) (3.45,0.65) (3.8,0.32) (4.8,0.16)};
\draw[thick,red] plot[smooth] coordinates {(0.4,3.2) (1.0,2.4) (1.6,1.6) (2.0,0.8) (2.5,0.4) (3.0,0.2) (3.5,0.1) (4.0,0.05)};
\fill[blue] (2.0,1.6) circle (2pt) node[above right] {$4\%$};
\fill[red] (2.0,0.8) circle (2pt) node[below right] {$1\%$};
\fill[blue] (3.45,0.65) circle (2pt);
\end{tikzpicture}
\caption{How to read the two budget calculations used in this work: The horizontal axis is a candidate RB budget $N$. The vertical axis
is overload probability: the probability that total footprint demand exceeds $N$. The red curve is the Monte Carlo estimate of that
probability. It crosses the target $\epsilon=1\%$ at $1000$ RBs, so $1000$ RBs is the Monte Carlo required budget in this schematic. The
blue curve is the conservative concentration calculation. For each sampled attenuation pattern, the concentration inequality gives an
upper bound on overload probability; the blue curve shows the average of those upper bounds across sampled attenuation patterns. This
blue curve crosses the same target at $1020$ RBs, so the concentration budget is $2\%$ higher than the Monte Carlo budget. Notice that
the two comparisons are different: the concentration budget is only $2\%$ higher, but at the Monte Carlo budget of $1000$ RBs the
concentration upper bound on overload probability is $4\%$. Thus a small budget gap does not necessarily mean that the concentration
bound at the Monte Carlo budget is close to the target probability.}
\label{fig:family_gap_vs_bound}
\end{figure}

\section{Background and Related Work}
\label{sec:related_work}

Prior work is examined with respect to the pre-deployment dimensioning question central to this study. Rather than assessing the general
utility of existing approaches, the analysis concentrates on their input assumptions and their ability to
generate reliable footprint-level RB budgets in the presence of spatially correlated attenuation.

We identify four distinct but related strands of prior work in the existing literature on dimensioning: attenuation modeling,
terrestrial RB dimensioning, spatial-correlation analysis, and satellite rain-fade resource dimensioning. Works which consider
traffic patterns and shadowing effects~\cite{bonald2003wireless,gomez2018stochastic} provide a foundation but
do not fully address the specific requirements of satellite footprint dimensioning: namely, the need for footprint-level
RB budget outputs under spatially correlated attenuation conditions.

\subsection{Propagation models describe attenuation, but not RB budgets}
\label{ssec:related_work_propagation}

Propagation recommendations provide an essential starting point because they describe how the radio channel can be attenuated by clutter,
rain, or other environmental effects. For example, ITU-R recommendations give statistically grounded models for terrestrial clutter loss
and Earth-space propagation impairments~\cite{iturp2108,iturp618}. Such models are directly relevant to a satellite operator because they
describe the physical loss process that must be accounted for before deployment.

However, an attenuation recommendation is not yet a footprint-dimensioning rule. A marginal attenuation model describes the distribution of
loss on one link. A footprint-level RB budget also requires a spatial model saying how losses at different user locations move together, a
traffic or user-location model, a link-budget-to-RB conversion, and an overload target. Here, the attenuation model is therefore
used as one component of a larger dimensioning pipeline rather than as the final output.

\subsection{Terrestrial RB dimensioning gives the right output type in a different system}
\label{ssec:related_work_terrestrial_rb}

Liu et al.~\cite{liu2021uplink} are the closest precedent in terms of output type. They study how many RBs are required in a terrestrial
cellular uplink under shadowing and an outage target. This is important because it connects a random spatial network model, shadowing,
and an overload criterion to a required number of RBs. In that sense, it addresses the same kind of operational object as the present
paper: a resource-block budget rather than only a link margin or an attenuation distribution.

The cellular machinery behind that RB budget is nevertheless specific to terrestrial uplink. Users transmit from the ground to base
stations on Earth. Their RB classes are generated through cellular association, scheduling, uplink interference, and the access scheme,
such as OMA or NOMA.\footnote{OMA, or orthogonal multiple access, separates users into different time-frequency resources. NOMA, or
non-orthogonal multiple access, lets users share a resource and separates them by signal processing. Cellular association is the rule by
which a user attaches to a serving base station. Scheduling is the rule by which the network chooses which attached users transmit on the
considered time-frequency resources. Uplink interference is unwanted received power created by other users transmitting at the same time.
The access scheme is the rule that decides whether users are separated onto different resources or allowed to share one resource.}
As demonstrated in stochastic-geometry analyses like Haenggi's study of active-user point processes~\cite{haenggi2017user}, this
network infrastructure determines how user activity translates into RB classes: base station locations, cellular association rules,
and scheduling decisions collectively assign RB classes to active users.

By contrast, the satellite-footprint problem makes spatial correlation an explicit dimensioning input rather than an incidental network
effect. Instead of deriving RB classes from cellular infrastructure elements (association rules, scheduling
decisions, and interference management), each user's RB allocation is computed directly from the value of a common spatial attenuation
field sampled at that user's location. This approach creates a fundamentally different correlation structure: spatial correlations in
RB demand are directly imposed by the attenuation field model. Without such explicit modeling, one would have essentially spatially independent
RB allocations (though spatial correlations in RB demand may still emerge as incidental byproducts of network-level resource allocation decisions).

The fundamental issue is that terrestrial RB-dimensioning formulas are built around network infrastructure assumptions that do not take into
account the spatial correlation structure of attenuation fields expected in satellite contexts. These formulas:
\begin{enumerate}
\item account more for obstacles in the terrestrial environment rather than the atmospheric effects that are more relevant for satellite links, and
\item preserve terrestrial mechanisms like cellular association and interference management that are very different in satellite contexts, and/or
\item ignore the spatial correlation structure by treating attenuation as independent across links
\end{enumerate}

Our contribution is to maintain the RB-budget objective while adapting it to satellite footprints where spatial covariance of attenuation
is a primary dimensioning input rather than a secondary propagation detail.

\subsection{Spatial correlation models describe propagation patterns, not footprint capacity requirements}
\label{ssec:related_work_spatial_correlation}
Kimura~\cite{kimura2023interference} studies spatially correlated shadowing within a stochastic-geometry framework,\footnote{A stochastic-geometry
framework models network elements, such as users or transmitters, as random spatial point patterns and studies the resulting distribution
of network quantities such as interference, coverage, or outage.} analyzing its impact
on interference statistics in non-Poisson networks.\footnote{A non-Poisson network is a random spatial network whose point locations are
not modeled as a homogeneous Poisson point process. This matters because repulsion, clustering, or other dependence between point
locations can change aggregate quantities such as interference.}

In doing so,
Kimura notes that Gudmundson's exponential correlation model~\cite{gudmundson1991correlation} is a classical way to express the empirical
idea that nearby shadowing values are more strongly related than distant ones:
\[
\rho(d)=\exp(-d/d_0),
\]
where $d$ is the distance between two locations and $d_0$ is the decorrelation distance.\footnote{The decorrelation distance is the
distance scale over which shadowing values stop moving strongly together. In the exponential model, larger $d_0$ means correlation decays
more slowly with distance.} Our simulations use the closely related Gaussian-kernel covariance in
\eqref{eq:rbf_corr_rule}, namely a rule proportional to $\exp(-d^2/(2\ell^2))$ rather than $\exp(-d/d_0)$. Here $\ell$ plays the same
role as $d_0$: it is the length scale controlling how quickly correlation falls with distance. Gudmundson's exponential model gives the
propagation basis for this choice: shadowing correlation is spatial and decays with separation distance.

The Gaussian kernel replaces linear distance in the exponent by squared distance, so the correlation falls more smoothly near the origin
and faster at long range. That is a modeling choice about the shape of the decay curve, not a contradiction of the Gudmundson principle.
Both kernels say that nearby users see more similar attenuation than distant users; the Gaussian kernel is additionally convenient because
it gives a consistent covariance matrix for every finite set of user locations sampled in a Monte Carlo trial.\footnote{In each trial, the
user locations change, so the simulation must rebuild a covariance matrix from the distances among that trial's users. The mean attenuation
level is fixed separately by the scenario. The covariance matrix controls how the sampled attenuation values vary together around that
mean. For this multivariate Gaussian draw to be well defined, the covariance matrix must be positive definite, meaning that every weighted
sum of the sampled field values has nonnegative variance. The standard Gaussian, or radial-basis-function, kernel has this property for
any finite set of locations. This is the practical reason for using the squared-distance form here. The denominator $2\ell^2$ is the usual
Gaussian-kernel normalization: at distance $d=\ell$, the correlation is $\exp(-1/2)\approx 0.61$.}

Kimura's own contribution is different from ours. Kimura studies how spatial correlation in both node locations\footnote{Here, a node is a
network point such as a transmitter, receiver, base station, or user equipment, depending on the model.} and shadowing affects
interference in non-Poisson networks. The output is an interference analysis: statistical expressions describing received interference
behavior under correlated propagation and correlated node placement. This is valuable because it shows that spatial correlation can change
wireless-network statistics in a substantial way. However, the interference analysis is more useful for modeling coverage quality than for
capacity dimensioning. It is thus complementary to the present work, but different in focus.

The present work uses spatial correlation for a different engineering question. We ask how a satellite operator should choose a footprint
RB budget for randomly located users when attenuation is spatially correlated. The output is therefore not an interference variance,
correlation coefficient, or propagation statistic. It is a footprint-level RB dimensioning rule: a method for translating a spatial
attenuation model and a user-location model into an RB budget that meets an overload-probability target.

\subsection{Satellite rain-fade dimensioning is closest, but targets different resources}
\label{ssec:related_work_rain_fade}

Lacoste et al.~\cite{lacoste2024resource} are the closest satellite-dimensioning neighbor. They study resource dimensioning for broadband
satellite return networks under spatially correlated rain fade. Their paper is especially useful because it explicitly analyzes how
satellite rain-fade dimensioning can be distorted by simplifying the dependence structure. In their setting, assuming that all terminals
fade together is too conservative and can overestimate required resources, while assuming independent fades can underestimate resources and
risk service-level violations. This analysis in the rain-fade return-link setting supports the broader point
that spatial dependence is not a secondary detail in satellite resource dimensioning.

Their work differs in the resource being dimensioned. Lacoste et al. dimension return-link
bandwidth and terminal-side transmit power for ground terminals transmitting back to the satellite.\footnote{A return link is the
direction from the ground terminal to the satellite, so the ground terminal is the transmitter.} Our output is the number of RBs available
for a satellite footprint. Our model starts from random user locations and a correlated attenuation field, converts each sampled
user-location attenuation value into capped RB demand, and then dimensions the aggregate footprint budget. While their work shows why spatial
correlation-aware satellite dimensioning is necessary, ours provides an RB dimensioning rule for a satellite footprint.

\subsection{Standards-based user-density reference points}
\label{ssec:related_work_density_reference}

The user intensity in this work is written as an areal density $\lambda$ in users per square metre. Telecom standards often use a
different convention. In the Dense Urban-eMBB evaluation configuration, ITU-R M.2412-0 gives a baseline user density of $10$ UEs per
TRxP\footnote{A TRxP is network infrastructure: a transmission and reception point, such as a base-station radio point in a terrestrial
cellular deployment. A UE is the opposite side of the link: user equipment, i.e., the terminal or device being served. eMBB means enhanced
mobile broadband, the high-throughput broadband service category in 5G evaluations.}~\cite{iturm2412}. The corresponding 3GPP
dense-urban scenario also uses $10$ users
per TRxP as the full-buffer baseline and notes that $20$ users per macro TRxP is not precluded~\cite{tr38913}. The associated
Dense Urban-eMBB user-experienced data-rate targets in ITU-R M.2410-0 are $100\,\mathrm{Mbit/s}$ downlink and $50\,\mathrm{Mbit/s}$
uplink~\cite{iturm2410}.

To compare these standards values with the PPP intensity used in our simulations, an area must be assigned to one TRxP. We use the
Dense Urban macro-layer inter-site distance (ISD), with $\mathrm{ISD}=200\,\mathrm{m}$ from the same 3GPP
scenario~\cite{tr38913} and approximate one TRxP service
region by a hexagonal area
\[
A_{\mathrm{TRxP}}=\frac{\sqrt{3}}{2}\mathrm{ISD}^2\approx 3.46\times 10^4\,\mathrm{m}^2.
\]
This gives
\begin{align*}
\frac{10}{A_{\mathrm{TRxP}}}
&\approx 2.9\times 10^{-4}\ \mathrm{users/m^2},\\
\frac{20}{A_{\mathrm{TRxP}}}
&\approx 5.8\times 10^{-4}\ \mathrm{users/m^2}.
\end{align*}
These values are not used as exact satellite-footprint measurements. They provide standards-based reference points for interpreting the
range of user densities considered in the footprint simulations.
In the satellite service case studied here, we use a much lower baseline throughput target of $1\,\mathrm{Mbps}$ total per user. A
reasonable footprint-density study therefore spans several orders of magnitude, from occasional-use densities around
$10^{-6}\,\mathrm{users/m^2}$ to heavy IoT-type densities around $10^{-3}\,\mathrm{users/m^2}$.

\section{System Model and RB Demand}
\label{sec:system_model_rb_demand}

We consider a satellite ground footprint $\mathcal{B}\subset\mathbb{R}^2$ served by a satellite RB pool.\footnote{A deployed satellite may
realize this footprint through multiple beams. In this work the dimensioned object is the footprint-level aggregate RB pool serving
$\mathcal{B}$, so the footprint is the common spatial region on which user locations and attenuation are modeled.} A user location is denoted by
$x\in\mathcal{B}$, where $x$ is a two-dimensional ground position inside that footprint. The footprint resource is counted in resource
blocks (RBs).\footnote{An RB (resource block) is a small time--frequency resource unit. In the numerical study, one RB has bandwidth
$W_{\mathrm{RB}}=180\,\mathrm{kHz}$, corresponding to the common $12\times 15\,\mathrm{kHz}$ PRB scale.} Each user has a target rate
$c$ in bits/s. The dimensioning problem is to choose a total footprint budget, measured in RBs, large enough to serve the aggregate demand with
the desired reliability.

For a user at location $x$, the rate carried by one RB is modeled through the Shannon--Hartley expression~\cite{shannon1948},
\begin{equation}
\label{eq:shannon_hartley_rb}
\begin{aligned}
R_{\text{RB}}(x)
&=W_{\mathrm{RB}}\,\eta(x),\\
\eta(x)
&:=\log_2\!\bigl(1+\mathrm{SNR}(x)\bigr).
\end{aligned}
\end{equation}
Here $W_{\mathrm{RB}}$ is the bandwidth of one RB and $\mathrm{SNR}(x)$ is the received signal-to-noise ratio at location
$x$. In a more general interference-aware treatment, the same expression would use the signal-to-interference-plus-noise ratio
\begin{equation}
\label{eq:sinr_def}
\mathrm{SINR}(x)=\frac{P_{\mathrm{rx}}(x)}{I(x)+N},
\end{equation}
where $I(x)$ is cochannel interference power and $N$ is noise power on one RB. We use the noise-limited specialization,
so interference is not modeled and the rate expression is written in terms of $\mathrm{SNR}(x)$. The number of RBs required by this user is therefore
\begin{equation}
\label{eq:nrb_demand}
N_{\mathrm{RB}}(x)
= \left\lceil \frac{c}{R_{\text{RB}}(x)} \right\rceil
= \left\lceil \frac{c}{W_{\mathrm{RB}}\log_2\!\bigl(1+\mathrm{SNR}(x)\bigr)} \right\rceil.
\end{equation}
This is the link-budget-to-RB mapping used throughout this work.

The formulas above reduce RB demand to a function of $\mathrm{SNR}(x)$. Following the original link-budget decomposition, we write
\begin{equation}
\label{eq:snr_pathloss}
\mathrm{SNR}(x) = K(x)\,d^{-\gamma},
\end{equation}
where $d$ is the distance to the satellite and $\gamma\approx 2$ is the free-space pathloss exponent. The location-dependent gain is
split into a fixed link-budget term and a random attenuation term,
\begin{equation}
\label{eq:kx_decomp}
K(x)=K_0S(x),
\end{equation}
where $K_0$ collects deterministic quantities such as transmit power, antenna gains, and noise level, while $S(x)$ is the random
shadowing factor.\footnote{In this convention, $S(x)=1$ means no shadowing loss, $S(x)<1$ means attenuation, and very small $S(x)$ means
deep shadowing.} Thus
\begin{equation}
\label{eq:snr_shadowing}
\mathrm{SNR}(x)=K_0S(x)d^{-\gamma}.
\end{equation}
Substituting \eqref{eq:snr_shadowing} into \eqref{eq:nrb_demand} makes each user's RB demand a function of the sampled shadowing value at
that user's location:
\begin{equation}
\label{eq:nrb_shadowing}
N_{\mathrm{RB}}(x)
= \left\lceil
\frac{c}{W_{\mathrm{RB}}\log_2\!\bigl(1+K_0 S(x)d^{-\gamma}\bigr)}
\right\rceil.
\end{equation}

Finally, the operator may impose a maximum counted demand of $M$ RBs per user. This cap is an operational policy: users whose raw value
of $N_{\mathrm{RB}}(x)$ exceeds $M$ are still counted, but their contribution to the footprint load is limited to
$\min\{N_{\mathrm{RB}}(x),M\}$ RBs. The total footprint demand in a trial is therefore obtained by summing these capped per-user
contributions.

\section{Gaussian Attenuation Model}
\label{sec:gaussian_attenuation_model}

The random attenuation term in Section~\ref{sec:system_model_rb_demand} is the shadowing factor $S(x)$. Because $S(x)$ is multiplicative
in linear power units, we model it on the log scale: multiplicative attenuation factors in linear units become additive offsets after
taking logarithms. Let
\begin{equation}
\label{eq:log_shadowing_def}
G(x)=\ln S(x)
\end{equation}
denote log-shadowing. Equivalently,
\begin{equation}
\label{eq:shadowing_exp}
S(x)=\exp\!\bigl(G(x)\bigr).
\end{equation}
Thus a Gaussian model for $G(x)$ corresponds to a lognormal model for the positive shadowing factor $S(x)$.

We model $G$ as a Gaussian random field over the footprint. This means that for any finite set of user locations
$x_1,\dots,x_n\in\mathcal{B}$, the vector
\[
(G(x_1),\dots,G(x_n))
\]
is multivariate Gaussian. The field is specified by a mean function $m(x)=\mathbb{E}[G(x)]$ and covariance function
$C(x,y)=\mathrm{Cov}(G(x),G(y))$. We use this finite-dimensional definition directly in the simulations: once a trial has
realized user locations $x_1,\dots,x_n$, we sample the Gaussian vector only at those locations rather than materializing a full
footprint-scale field map.

For a finite set of locations, write
\[
\Sigma_{ij}:=\mathrm{Cov}(G(x_i),G(x_j)),\qquad i,j=1,\dots,n.
\]
We assume a common marginal variance
\begin{equation}
\label{eq:equal_marginal_variance}
\mathrm{Var}(G(x_1))=\mathrm{Var}(G(x_2))=\cdots=\mathrm{Var}(G(x_n))=\sigma^2,
\end{equation}
and use the standard isotropic radial basis function (RBF) correlation rule
\begin{equation}
\label{eq:rbf_corr_rule}
\mathrm{Corr}(G(x_i),G(x_j))=\exp\!\left(-\frac{\|x_i-x_j\|_2^2}{2\ell^2}\right),\qquad i\neq j,
\end{equation}
where $\ell>0$ is the correlation length.\footnote{There are three separate uses of ``2'' in this expression. The subscript in
$\|\cdot\|_2$ means Euclidean distance: for ground positions $x=(x_1,x_2)$ and $y=(y_1,y_2)$, $\|x-y\|_2$ is the ordinary straight-line
distance $\sqrt{(x_1-y_1)^2+(x_2-y_2)^2}$. The superscript in $\|x_i-x_j\|_2^2$ squares that distance before putting it into the
exponential, giving the smooth RBF decay. The scale $\ell$ says how quickly correlation falls: at distance $\ell$, the correlation is
$\exp(-1/2)\approx 0.61$, while at distance $2\ell$ it is $\exp(-2)\approx 0.14$. The separate factor $2$ in the denominator $2\ell^2$
is only a length-scale convention.} Classical shadow-fading models also use distance-based correlation rules, notably
Gudmundson's exponential model~\cite{gudmundson1991correlation}. We use the RBF rule here as a controlled positive-definite
covariance model with one interpretable length scale.\footnote{Positive definite means that, for any finite set of user locations, the
covariance matrix assigns a nonnegative variance to every weighted sum of sampled field values. This is required for the multivariate
Gaussian vector sampled in the simulation to be well defined.}

For example, with three locations, define $d_{ij}:=\|x_i-x_j\|_2$ and
\[
r_{ij}:=\exp\!\left(-\frac{d_{ij}^2}{2\ell^2}\right).
\]
The covariance matrix is then
\[
\Sigma =
\sigma^2
\begin{bmatrix}
1 & r_{12} & r_{13}\\
r_{12} & 1 & r_{23}\\
r_{13} & r_{23} & 1
\end{bmatrix},
\qquad 0\le r_{ij}\le 1.
\]
This matrix is defined for the log-shadowing values $G(x_i)$, while the corresponding linear shadowing factors are obtained through
$S(x_i)=\exp(G(x_i))$.

\subsection{Radial RB classes and field-induced RB classes}
\label{ssec:radial_to_field_classes}

\begin{figure}[t]
\centering
\includegraphics[width=0.68\linewidth]{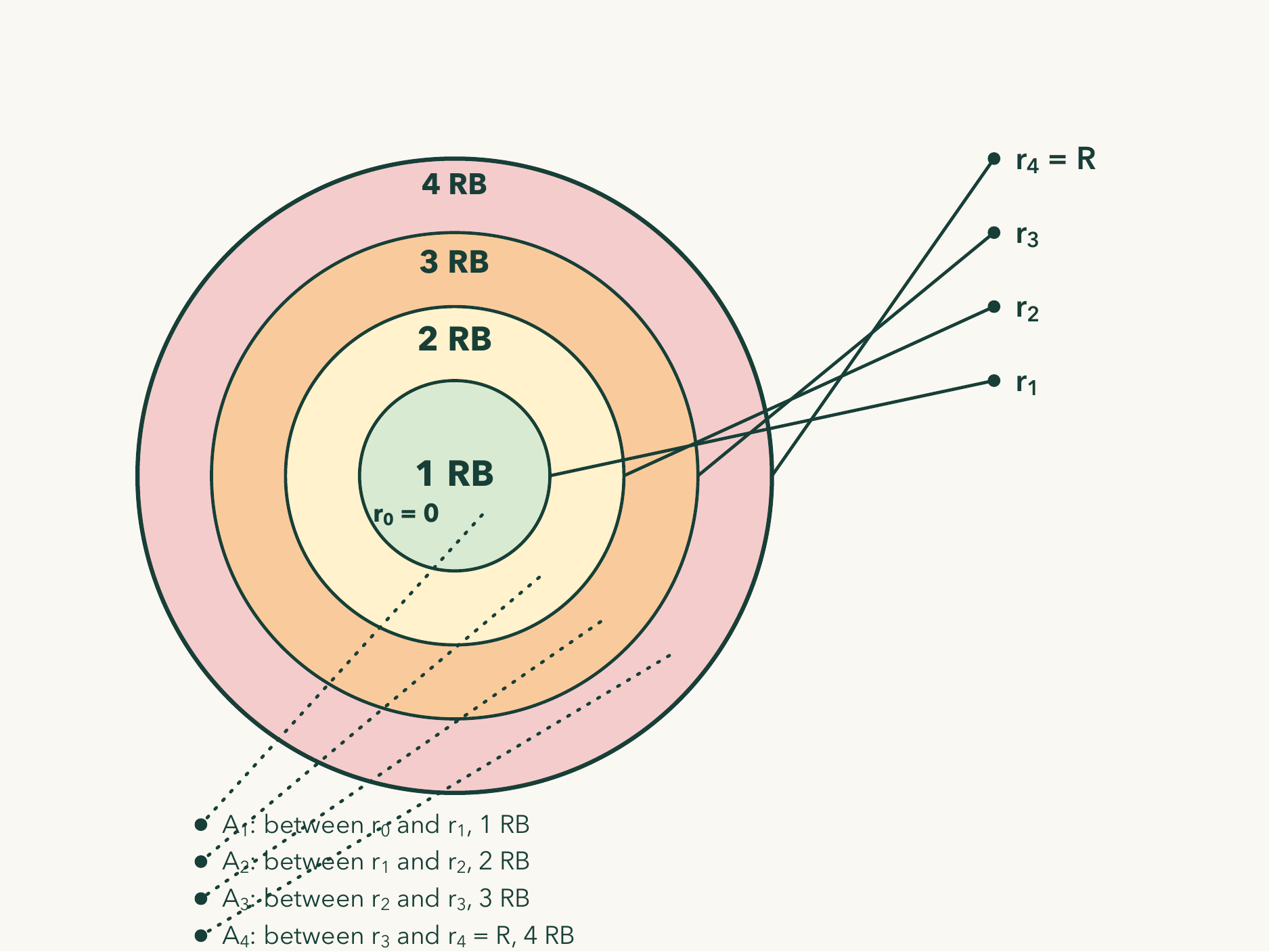}
\caption{Deterministic radial RB classes. In the simplified radial picture, users are grouped into concentric annuli according to the
number of RBs required at their distance from the footprint center.}
\label{fig:rb_annuli}
\end{figure}

It is useful to first consider a deterministic radial picture. If all location dependence were summarized by distance from the footprint
center, the per-user RB demand could be written as a function of radius,
\[
N_{\mathrm{RB}}^{\mathrm{rad}}(x)=\left\lceil\frac{c}{W_{\mathrm{RB}}\,\log_2\!\left(1+\frac{K}{\|x\|^\gamma}\right)}\right\rceil,
\]
where $K$ is a channel constant in this simplified radial formula. This thought experiment creates concentric RB classes: for thresholds
\[
0=r_0<r_1<\cdots<r_{K_{\mathrm{det}}}=R,
\]
one has
\[
N_{\mathrm{RB}}^{\mathrm{rad}}(x)=k
\quad\Longleftrightarrow\quad
r_{k-1}\le \|x\|<r_k.
\]
The corresponding annulus is
\[
A_k:=B(0,r_k)\setminus B(0,r_{k-1}),
\qquad
\zeta_k:=\Phi(A_k),
\]
and the deterministic radial total demand is
\[
N_{\mathrm{total}}=\sum_{k=1}^{K_{\mathrm{det}}} k\,\zeta_k.
\]
Figure~\ref{fig:rb_annuli} shows the geometric idea behind this decomposition. The important point is not the circular shape itself; it
is that total demand can be counted by first assigning each location to an RB class and then counting how many users fall in each class.

\begin{figure}[t]
\centering
\includegraphics[width=0.82\linewidth]{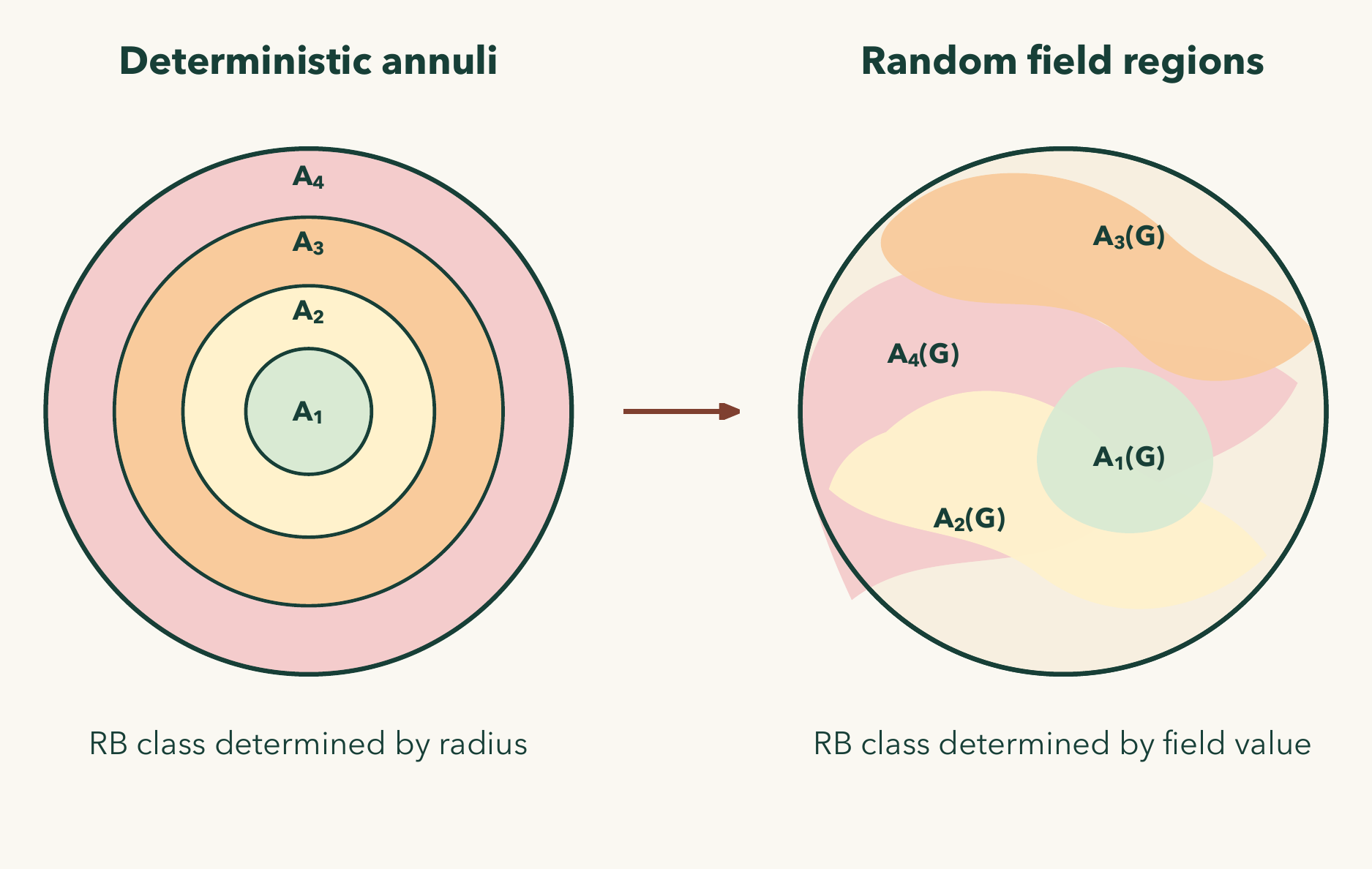}
\caption{Field-induced RB regions under spatial attenuation. Compared with the circular classes in Figure~\ref{fig:rb_annuli}, a
spatial attenuation field produces generally non-circular subsets $A_k(G)$ of the footprint.}
\label{fig:annuli_to_field_regions}
\end{figure}

The Gaussian-field model replaces these concentric annuli by field-induced regions. Once a log-attenuation field realization $G$ is
fixed, the RB demand is no longer determined only by radius. A location belongs to the $k$-RB class if its field-induced demand equals
$k$:
\[
A_k(G):=\{x\in\mathcal{B}:N_{\mathrm{RB}}(x;G)=k\}\subseteq\mathcal{B}.
\]
These sets need not be circular, because $G(x)$ need not vary radially. Given a user configuration $\Phi$, the corresponding class count
is
\[
\zeta_k(G,\Phi):=\Phi(A_k(G)).
\]
Figure~\ref{fig:annuli_to_field_regions} shows what changes when the RB classes are induced by a spatial attenuation field rather than
by radius alone. The reader may compare it with Figure~\ref{fig:rb_annuli}: the same class-counting logic remains, but the regions
that generate those counts are now irregular and field-dependent.

Thus the annulus picture should be read as a geometric thought experiment: it explains why total demand can be grouped by RB class, while
the Gaussian-field model supplies the actual, generally amorphous, class regions. In the Monte Carlo study, these regions are
not materialized as full footprint-scale maps. Instead, the sampled Gaussian vector at the realized user locations directly labels the
users by their capped RB demands.

\section{Gaussian Scenario Setup}
\label{sec:gaussian_scenario_setup}

The empirical study varies only the Gaussian attenuation family. A scenario is denoted by
\[
\mathrm{G}(\theta),
\qquad
\theta=(m,\sigma^2,\ell),
\]
where $m$ is the constant mean of the log-shadowing field $G(x)$, $\sigma^2$ is its marginal variance, and $\ell$ is the RBF
correlation length from \eqref{eq:rbf_corr_rule}. Within one scenario, these three quantities are fixed for all Monte Carlo trials.
Trial-to-trial variation comes from the random user locations and from the sampled Gaussian vector at those realized locations.

To model user locations, we need a way to draw both how many users are present and where they are inside the footprint. A Poisson point
process (PPP) does this in a simple way: users are placed randomly over the footprint, the number of users in any chosen region follows a
Poisson distribution,\footnote{Intuitively, a Poisson distribution is the standard model for counting independent random arrivals in a
fixed region or time interval. As the region or time interval becomes smaller, the expected number of arrivals goes to zero, while larger
regions or intervals have proportionally larger expected counts. If the expected count is $\mu$, then
$\mathbb{P}(N=n)=e^{-\mu}\mu^n/n!$ for $n=0,1,2,\dots$.} and counts in non-overlapping regions are
independent~\cite{baccelli2009stochastic}.

We therefore use a PPP for the random user field. The PPP assumption is also useful beyond simulation. We also want a conservative way
to upper bound the probability that the total RB demand is much larger than its average value by some amount --- an inequality that tells
us how probable it is that the random demand deviates from the average by a given amount ---  a concentration inequality. The concentration inequality
we use is designed for sums formed from Poisson-distributed points~\cite{decreusefond2012robust}. That matches our setting: the users are the
Poisson-distributed points, each user contributes an RB demand, and the total footprint demand is the sum of those contributions.

The fixed experimental controls are the same across all scenarios. The footprint is a disk of radius $20\,\mathrm{km}$, users are
drawn from a PPP with intensity $\lambda=10^{-5}$ users per square metre, and the overload target is $\epsilon=0.01$. Each user has target rate
$c=1\,\mathrm{Mbit/s}$, each RB has bandwidth $W_{\mathrm{RB}}=180\,\mathrm{kHz}$, the reference SNR prefactor in
\eqref{eq:snr_shadowing} is set to one, and the operator cap is $M=20$ RBs per user. The Monte Carlo study uses $T=100$ trials per
scenario.

The six Gaussian scenarios combine three mean log-shadowing levels with two correlation lengths:
\begin{align*}
m\in\{-1.5,-0.5,-0.25\},\qquad
\sigma^2=0.2,\\
\ell\in\{500\,\mathrm{m},20\,\mathrm{km}\}.
\end{align*}
The mean values represent stronger, medium, and mild average attenuation. The two length scales compare short-range correlation with
correlation at footprint scale while keeping the marginal variance fixed. This lets us isolate how spatial correlation changes the
footprint-level RB budget, rather than mixing that effect with a change in one-point attenuation variability.

\section{Monte Carlo Dimensioning}
\label{sec:monte_carlo_dimensioning}

For each Gaussian scenario $\mathrm{G}(\theta)$, the Monte Carlo procedure produces a sample of total footprint demands. In trial
$t=1,\dots,T$, users are drawn as a PPP realization $\Phi_t$ on the footprint. Conditional on those realized user locations, the
Gaussian vector
\[
g_t=\bigl(G(x):x\in\Phi_t\bigr)
\]
is sampled from the scenario's multivariate normal distribution. This sampled vector is the realized representation of the field used in the simulation
in that trial; no full footprint-scale field map is constructed.

Each sampled value is converted to a shadowing factor through $S(x)=\exp(G(x))$, then to a raw per-user RB demand using
\eqref{eq:nrb_shadowing}, and finally to a capped contribution $\min\{N_{\mathrm{RB}}(x),M\}$. This gives one total footprint demand
$D_t^{\mathrm{G}(\theta)}$ for the trial. This is the quantity whose upper quantile determines the footprint-level dimensioning budget.

Equivalently, for active user locations $x_1,\dots,x_n$, let $\mathcal{U}_k$ be the set of users whose capped contribution is exactly
$k$ RBs ($k=1,\dots,N_{\mathrm{RB}}^{\max}$), where $N_{\mathrm{RB}}^{\max}=M$ in the capped model used here. Users with raw
demands above $M$ therefore enter the $M$-RB class rather than being removed from the load calculation. This
gives a decomposition of total demand by RB classes:
\begin{equation}
\label{eq:ntotal_class_decomp}
N_{\mathrm{total}}
=
\sum_{k=1}^{N_{\mathrm{RB}}^{\max}} k\,|\mathcal{U}_k|,
\end{equation}
and the overload event is
\begin{equation}
\label{eq:ntotal_overload_class}
\left\{N_{\mathrm{total}}>N_{\mathrm{avail}}\right\}
=
\left\{\sum_{k=1}^{N_{\mathrm{RB}}^{\max}} k\,|\mathcal{U}_k|>N_{\mathrm{avail}}\right\}.
\end{equation}

For an integer footprint budget $N$, define the trial overload indicator
\[
I_t^{\mathrm{G}(\theta)}(N)
:=
\mathbf{1}\{D_t^{\mathrm{G}(\theta)}>N\}.
\]
The scenario-specific overload probability is
\begin{equation}
\label{eq:pout_conditional_definition}
p_{\mathrm{out}}^{\mathrm{G}(\theta)}(N)
:=
\mathbb{P}\!\left(D_t^{\mathrm{G}(\theta)}>N\right).
\end{equation}
Its Monte Carlo estimator is
\begin{equation}
\label{eq:pout_mc_estimator}
\widehat{p}_{\mathrm{out}}^{\mathrm{G}(\theta)}(N)
:=
\frac{1}{T}\sum_{t=1}^{T} I_t^{\mathrm{G}(\theta)}(N).
\end{equation}

The required budget is the smallest integer budget whose empirical overload probability is at most the target $\epsilon$:
$\widehat{N}_{\mathrm{req}}^{\mathrm{G}(\theta)}:=\min\{N\in\mathbb{Z}_{+}:
\widehat{p}_{\mathrm{out}}^{\mathrm{G}(\theta)}(N)\le\epsilon\}$.
Equivalently, we sort the trial totals
$D_1^{\mathrm{G}(\theta)},\dots,D_T^{\mathrm{G}(\theta)}$ and turn that ordered sample into a selector
\[
\epsilon\mapsto \widehat{Q}_{\mathrm{G}(\theta)}(\epsilon).
\]
In this notation,
\[
\widehat{N}_{\mathrm{req}}^{\mathrm{G}(\theta)}=\widehat{Q}_{\mathrm{G}(\theta)}(\epsilon).
\]
For the target $\epsilon=0.01$, this is the simulation-side 99th-percentile-style capped-load footprint budget for the scenario.
Because the study uses the same $T=100$ trials for every scenario, these empirical budgets should be read primarily as comparative
scenario outputs. The absolute upper-tail estimate is coarse at a $1\%$ target, but the comparison is controlled: footprint geometry,
user density, link-budget assumptions, cap, and overload target are held fixed while the Gaussian attenuation family is varied.

\section{Robust Upper Bound on Overload Probability}
\label{sec:concentration_comparator}

In addition to the Monte Carlo budget, we compute an averaged-bound concentration budget based on the Poisson-functional\footnote{A functional is
a quantity computed from an entire random object rather than from one scalar random variable. Here the random object is the Poisson user
configuration in the footprint, and the functional is the total RB demand induced by that configuration.} concentration approach developed
in~\cite{decreusefond2012robust}. Its role is not to replace the simulation study. For any candidate RB budget, it upper bounds the
probability that total footprint demand exceeds that budget.

The concentration calculation starts from one sampled realization inside a fixed Gaussian scenario $\mathrm{G}(\theta)$. For trial $i$, the PPP user
set is $\Phi_i$ and the sampled Gaussian vector at those user locations is $g_i$. As before, the simulation does not construct a full
footprint-scale field map; the vector $g_i$ is the realized field representation used in the calculation. The vector is converted into
capped per-user RB demands, and those demands are summarized by three quantities: $\mu_{G_i}$, $V_{G_i}$, and $M$.

Conceptually, if a fixed field representation $G_i$ were available over the whole footprint, it would define a capped demand function
\[
q_i(x)=\min\{N_{\mathrm{RB}}(x;G_i),M\}.
\]
The individual-realization mean $\mu_{G_i}$ is the Poisson mean of the total capped footprint demand, and $V_{G_i}$ is the corresponding
squared-demand scale:
\[
\mu_{G_i}=\lambda\int_{\mathcal{B}} q_i(x)\,dx,
\qquad
V_{G_i}=\lambda\int_{\mathcal{B}} q_i(x)^2\,dx.
\]
The implementation estimates these quantities from the same sampled Gaussian vector at the realized PPP user locations, rather than
from a stored footprint-scale map. If the trial contains $n_i$ users with capped demands $q_{i1},\dots,q_{in_i}$, the sampled vector is
used to estimate the average capped one-user demand and average squared capped one-user demand. These averages are then multiplied by
the expected number of users $\lambda|\mathcal{B}|$:
\[
\widehat{\mu}_{G_i}
=
\lambda|\mathcal{B}|\,\frac{1}{n_i}\sum_{j=1}^{n_i} q_{ij},
\qquad
\widehat{V}_{G_i}
=
\lambda|\mathcal{B}|\,\frac{1}{n_i}\sum_{j=1}^{n_i} q_{ij}^{2}.
\]
This is why the concentration calculation uses the sampled field representation: it supplies the user-location demand values from which
the individual-realization integral quantities are approximated. The quantity $M$ is the per-user RB-demand bound fed to the inequality. In the
numerical study, $M$ is the same operator cap used in the Monte Carlo simulation, namely
\[
M=20.
\]
Thus both the simulation and the concentration calculation work with capped total demand: a user whose raw RB demand exceeds $M$ is not dropped, but
contributes $M$ RBs to the total footprint load.

For one sampled field representation $G_i$, the individual-realization concentration bound is
\[
\mathbb{P}\!\left(N_{\mathrm{total}}\ge \mu_{G_i}+a \,\middle|\, G_i\right)
\le
\exp\!\bigl(-h_i(a)\bigr),
\qquad a\ge 0,
\]
where
\[
h_i(a)
=
\frac{V_{G_i}}{M^2}\,
g\!\left(\frac{aM}{V_{G_i}}\right),
\qquad
g(u)=(1+u)\ln(1+u)-u.
\]
This is a conditional bound attached to one sampled field representation. It is not yet a scenario-level budget.

To compare with the Monte Carlo output, we turn the individual-realization bounds into a scenario-level quantity by averaging them across the
sampled Gaussian field representations from the same scenario:
\[
\overline{B}_{\mathrm{G}(\theta)}(N)
=
\frac{1}{T}\sum_{i=1}^{T} B_i(N),
\]
where $B_i(N)$ is the individual-realization concentration upper bound on overload probability at budget $N$. For a candidate budget, the
individual-realization overload-probability bound is evaluated for each sampled realization in the scenario and then averaged across those sampled
realizations. The resulting averaged-bound quantity is then inverted in the same spirit as the Monte Carlo selector: the concentration
budget is the smallest budget whose averaged upper bound is below the target $\epsilon$.

\section{Results}
\label{sec:results}

Table~\ref{tab:results_summary} summarizes the six Gaussian scenarios and compares the simulation-derived budget
with the averaged-bound concentration budget for the same overload target. It should be read as a sensitivity analysis over the
Gaussian attenuation parameters, especially the mean attenuation level and the correlation length.

\begin{table*}[!t]
\centering
\caption{Scenario-wise Monte Carlo and concentration budgets. The target overload probability is $\epsilon=0.01$ and the per-user cap is $M=20$ RBs in every row. ``Avg-bound'' means that concentration bounds are first computed for individual sampled realizations and then averaged across the scenario. The averaged-bound gap is $(N_c^{\mathrm{avg}}-\widehat{N}_{\mathrm{req}}^{\mathrm{G}(\theta)})/\widehat{N}_{\mathrm{req}}^{\mathrm{G}(\theta)}$, comparing the averaged-bound concentration budget with the Monte Carlo budget. The range column uses the same percentage-gap calculation, $(N_{c,i}-\widehat{N}_{\mathrm{req}}^{\mathrm{G}(\theta)})/\widehat{N}_{\mathrm{req}}^{\mathrm{G}(\theta)}$, but for each individual sampled realization $i$; it reports the smallest and largest such gaps. A negative value means that the individual-realization concentration budget is below the Monte Carlo budget, while a positive value means it is above the Monte Carlo budget. The averaged overload-probability bound column fixes the Monte Carlo budget and reports the averaged concentration upper bound at that budget.}
\label{tab:results_summary}
\renewcommand{\arraystretch}{1.18}
\normalsize
\resizebox{0.96\textwidth}{!}{%
\begin{tabular}{c|l|r|r|r|r|c}
\hline
ID & Scenario & \shortstack{MC required\\budget} & \shortstack{Avg-bound\\conc. budget} & \shortstack{Gap:\\avg-bound conc.\\vs MC} & \shortstack{Avg overload-prob.\\upper bound\\at MC budget} & \shortstack{Range:\\individual conc.\\budgets vs MC} \\
\hline
A & $m=-1.50$, short-range corr. ($\ell=500$ m) & 223,983 & 226,056 & +0.9\% & 0.090 & [-1.0\%, +1.8\%] \\
B & $m=-1.50$, footprint-scale corr. ($\ell=20$ km) & 255,121 & 256,595 & +0.6\% & 0.039 & [-52.6\%, +1.2\%] \\
C & $m=-0.50$, short-range corr. ($\ell=500$ m) & 120,568 & 121,451 & +0.7\% & 0.042 & [-3.6\%, +2.1\%] \\
D & $m=-0.50$, footprint-scale corr. ($\ell=20$ km) & 183,031 & 214,443 & +17.2\% & 0.020 & [-66.5\%, +20.4\%] \\
E & $m=-0.25$, short-range corr. ($\ell=500$ m) & 100,134 & 100,927 & +0.8\% & 0.047 & [-3.3\%, +2.1\%] \\
F & $m=-0.25$, footprint-scale corr. ($\ell=20$ km) & 150,999 & 175,051 & +15.9\% & 0.020 & [-64.8\%, +19.1\%] \\
\hline
\end{tabular}}
\end{table*}

The main result is not merely that the correlation length changes the required budget. It is that, even with the marginal
log-attenuation variance fixed at $\sigma^2=0.2$, the induced RB-demand distribution changes substantially. Moving from short-range
correlation to footprint-scale correlation increases the simulation budget from $223{,}983$ to
$255{,}121$ RBs in the stronger-attenuation setting, from $120{,}568$ to $183{,}031$ RBs in the medium setting, and from $100{,}134$ to
$150{,}999$ RBs in the mild setting. These are increases of about $14\%$, $52\%$, and $51\%$, respectively. The effect is therefore not
explained by one-point attenuation statistics alone; it comes from how user demands move together across the footprint.

The demand spread shows the same point more directly. Under short-range correlation, the 99th percentile of total capped demand is only
about $5.2$k, $4.3$k, and $3.4$k RBs above the mean in the stronger, medium, and mild attenuation settings, respectively. Under
footprint-scale correlation, the corresponding gaps are about $40.2$k, $69.0$k, and $56.3$k RBs. Footprint-scale correlation therefore
introduces scenario-level fluctuations in which many users are simultaneously shifted toward better or worse attenuation conditions.

\begin{figure*}[!t]
\centering
\includegraphics[width=0.96\textwidth]{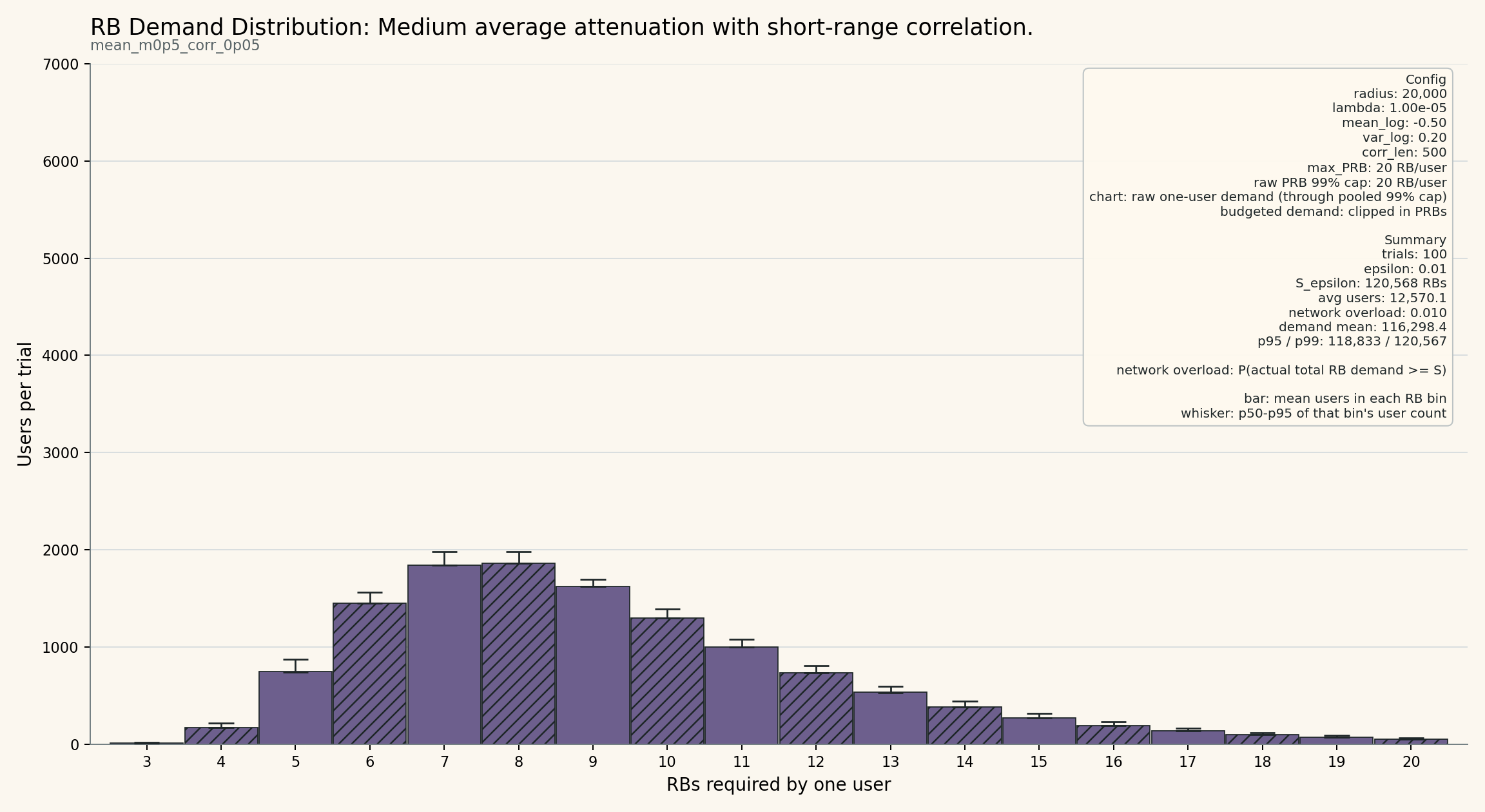}

\vspace{0.25em}

\includegraphics[width=0.96\textwidth]{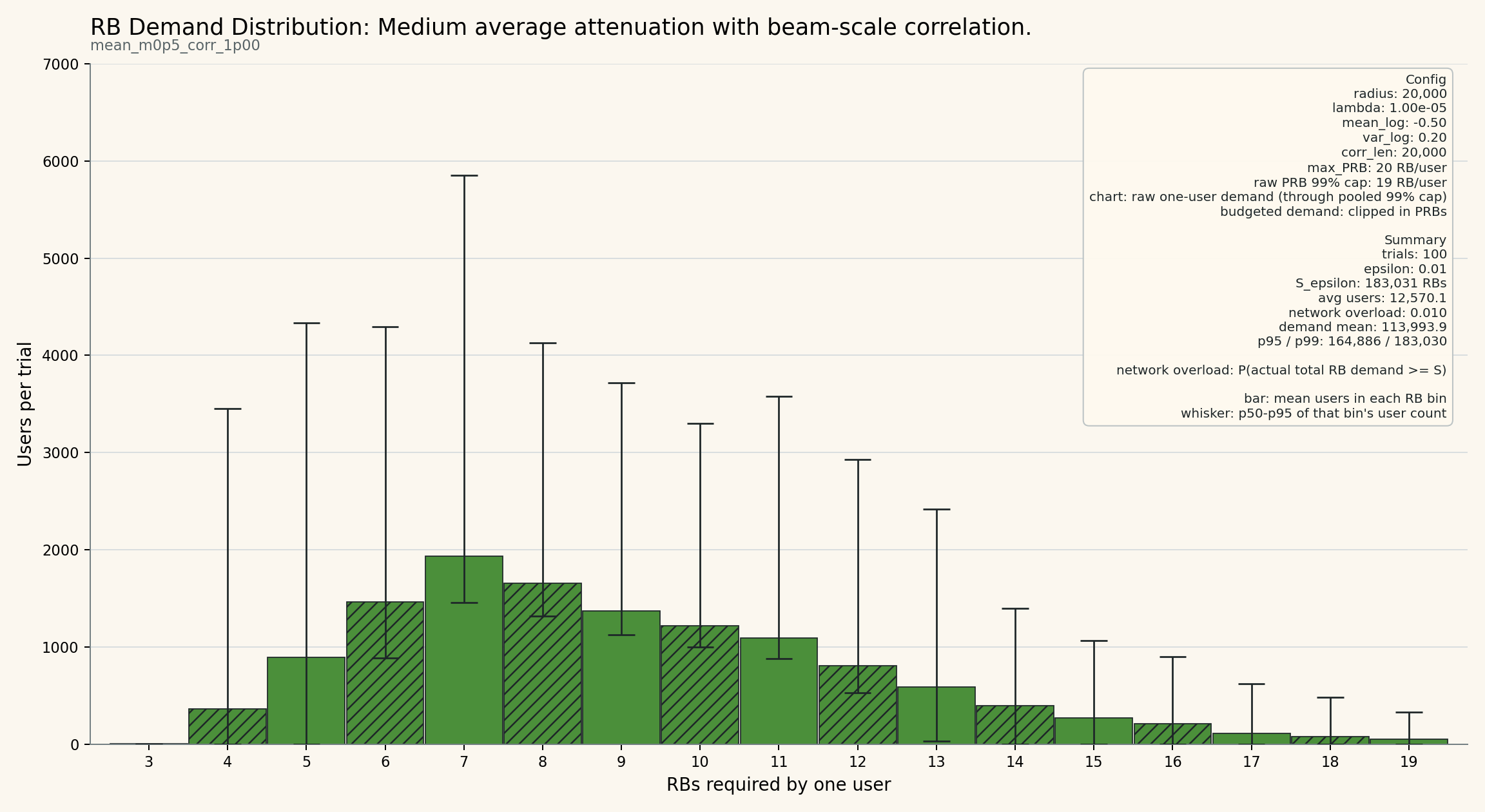}
\caption{Per-RB-bin demand distributions for the medium-attenuation scenario. Top: short-range correlation. Bottom: footprint-scale correlation.
Each bar gives the average number of users in that RB-demand bin across Monte Carlo trials. The whiskers show the
median-to-95th-percentile interval for that bin count across trials. The key effect is the much larger trial-to-trial variation in each
RB bin under footprint-scale correlation.}
\label{fig:medium_rb_bin_distribution}
\end{figure*}

The medium-attenuation case in Figure~\ref{fig:medium_rb_bin_distribution} shows this distributional change at the level of individual
RB classes. The average number of users in each RB bin is not the main visual difference. The striking change is the spread around those
bin counts. For example, in the $6$-RB bin, the median-to-95th-percentile interval is about $1{,}450$ to $1{,}550$ users under
short-range correlation, but about $850$ to $4{,}250$ users under footprint-scale correlation. In the $10$-RB bin, the same interval
widens from about $1{,}300$--$1{,}400$ users to about $1{,}000$--$3{,}300$ users. Thus footprint-scale correlation does not merely shift a
scalar budget. It changes how stable or unstable the whole RB-demand histogram is from one trial to the next.
When reading Figure~\ref{fig:medium_rb_bin_distribution}, the comparison to make is vertical rather than horizontal: for the same
medium-attenuation scenario, the top panel has relatively stable RB-bin counts, while the bottom panel has much taller whiskers across
many bins. Those whiskers show larger trial-to-trial variation within the RB-demand classes, which is consistent with the distributional
mechanism by which footprint-scale correlation can produce a larger aggregate budget.

The averaged-bound concentration budget is conservative in every scenario, but its behavior depends strongly on the correlation length.
The ``Range: individual conc. budgets vs MC'' column in Table~\ref{tab:results_summary} reports how far the most extreme
individual-realization concentration budgets are from the Monte Carlo budget. More precisely, each sampled field representation gives one
individual-realization concentration budget $S_c(G_i,\epsilon)$, and the table reports the smallest and largest percentage gaps between
those individual budgets and the Monte Carlo budget $\widehat{N}_{\mathrm{req}}^{\mathrm{G}(\theta)}$. A negative value means that an
individual-realization concentration budget is below the Monte Carlo budget; a positive value means it is above the Monte Carlo budget.
For the short-range cases, the averaged-bound concentration budget is within one percent of the Monte Carlo budget, and the individual-realization gaps are also small:
roughly $-4\%$ to $+2\%$ across the three short-range scenarios. For the footprint-scale cases, the averaged-bound concentration budget
remains above the Monte Carlo budget, but the individual-realization gaps become much wider. This is expected: one footprint-scale field
representation can be unusually favorable or unfavorable over a large part of the footprint, so the concentration budget computed from
that one representation can sit far from the Monte Carlo scenario budget.

For an operator, this range is informative in its own right. It gives a direct indication of how much the RB requirement can vary across
sampled attenuation representations within the same nominal scenario. In other words, the individual-realization range shows the scale of
scenario-internal RB-budget variability that the operator may need to manage.

This also explains why the correlation scale matters operationally. If the operator can estimate whether attenuation correlation is
short-range or footprint-scale, then the table indicates whether the concentration inequality is likely to give a robust scenario-level
budget from sampled realizations. In the short-range cases the individual-realization gaps are small, while footprint-scale correlation
produces much larger variation across realizations.

Operationally, this means that the covariance model is a dimensioning input. If the operator assumes only short-range correlation, the
required footprint budget is close to the average capped load plus a small safety margin. If the operator allows footprint-scale correlation,
the same one-point attenuation law can require a substantially larger budget because the aggregate load is more volatile. The
concentration calculation is useful in this setting because it expresses that volatility as a conservative upper bound on overload
probability for the same capped-load quantity estimated by Monte Carlo.

\section{Discussion}
\label{sec:discussion}

Let us consider the practical implications of the results. As a reasonable practical comparison point, suppose a satellite can make
$64$ beams available with $270$ RBs per beam, giving a total of $17{,}280$ RBs per satellite. Since the object dimensioned in this work is the footprint-level aggregate RB pool, we use this
$17{,}280$-RB value as an optimistic satellite-level comparison point. Under even this aggregate comparison, the Monte Carlo budgets in
Table~\ref{tab:results_summary} are infeasible for a single satellite.

This tells us that $1\,\mathrm{Mbps}$ internet over a $20\,\mathrm{km}$ radius footprint is not feasible with a single satellite at a user
density of $10^{-5}$ users per square metre. High-density user scenarios may approach $10^{-4}$ users per square metre, which would increase
the required RB budget by a factor of $10$, making the problem even more infeasible.

Our methodology may therefore be used to explore how the required RB budget changes with user density, target rate, footprint radius, and
other parameters.

The following use cases should be read as illustrative scaling estimates based on the studied short-range scenarios, not as
new link-budget simulations. In the radius calculation, the per-user capped RB-demand distribution is held fixed and only the
expected number of users is scaled through footprint area and density. In the IoT calculation, the $60\,\mathrm{kbps}$ classes are
obtained by conservatively rescaling the already-rounded $1\,\mathrm{Mbps}$ RB classes, rather than recomputing the Shannon expression
from scratch. We compare against the full $17{,}280$-RB satellite pool because the dimensioned object is the footprint-level aggregate RB
pool. The short-range concentration budgets are used below because, in Table~\ref{tab:results_summary}, they closely track the Monte
Carlo budgets with minimal deviation.

\subsection{Feasible footprint radius for 1 Mbps service}
\label{ssec:discussion_feasible_radius}

For this illustrative radius scaling, we condition on the PPP user model, fixed propagation assumptions, and short-range correlation. The
per-user capped RB-demand distribution is held fixed, so the first-order scaling comes from the expected number of users in the
footprint. We therefore scale the short-range $20\,\mathrm{km}$ concentration budget by user density and footprint area. If
$B_{\mathrm{G}(\theta)}^{1\,\mathrm{Mbps}}$ is the short-range concentration budget for scenario $\mathrm{G}(\theta)$ at radius $R_0=20\,\mathrm{km}$ and density
$\lambda_0=10^{-5}\,\mathrm{m}^{-2}$, then the scaled budget at radius $R$ and density $\lambda$ is approximated by
\begin{equation}
\label{eq:discussion_radius_budget_scaling}
B_{\mathrm{G}(\theta)}^{1\,\mathrm{Mbps}}(R,\lambda)
\approx
B_{\mathrm{G}(\theta)}^{1\,\mathrm{Mbps}}
\frac{\lambda \pi R^2}{\lambda_0 \pi R_0^2}
=
B_{\mathrm{G}(\theta)}^{1\,\mathrm{Mbps}}
\frac{\lambda R^2}{\lambda_0 R_0^2}.
\end{equation}
The satellite-level feasibility condition is $B_{\mathrm{G}(\theta)}^{1\,\mathrm{Mbps}}(R,\lambda)\le B_{\mathrm{sat}}$. Solving the equality case for
$R$ gives the feasible radius estimate
\begin{equation}
\label{eq:discussion_feasible_radius}
R_{\max}(\mathrm{G}(\theta),\lambda)
=
R_0
\sqrt{
\frac{B_{\mathrm{sat}}}{B_{\mathrm{G}(\theta)}^{1\,\mathrm{Mbps}}}
\frac{\lambda_0}{\lambda}
}.
\end{equation}
Table~\ref{tab:discussion_feasible_radius} gives the resulting approximate radii.
\begin{table}[!t]
\centering
\caption{Approximate feasible footprint radius \(R_{\max}\) for \(1\,\mathrm{Mbps}\) service under the short-range attenuation scenarios. The three right columns show \(R_{\max}\) in kilometres for user densities in \(\mathrm{m}^{-2}\).}
\label{tab:discussion_feasible_radius}
\small
\setlength{\tabcolsep}{2.5pt}
\begin{tabular}{lrrrr}
\hline
Attenuation & \(B^{1\,\mathrm{Mbps}}\) &
\(\lambda=10^{-6}\) &
\(\lambda=10^{-5}\) &
\(\lambda=10^{-4}\) \\
\hline
Stronger & \(226{,}056\) & \(17.5\) km & \(5.5\) km & \(1.7\) km \\
Medium   & \(121{,}451\) & \(23.9\) km & \(7.5\) km & \(2.4\) km \\
Mild     & \(100{,}927\) & \(26.2\) km & \(8.3\) km & \(2.6\) km \\
\hline
\end{tabular}
\end{table}
Thus a $20\,\mathrm{km}$ footprint with $1\,\mathrm{Mbps}$ per user is not close to feasible under the studied density
$10^{-5}\,\mathrm{m}^{-2}$. At the lower density $10^{-6}\,\mathrm{m}^{-2}$, the same satellite-level RB pool can support a
footprint-radius scale comparable to $20\,\mathrm{km}$ in the medium and mild attenuation settings. Increasing the density from
$10^{-5}$ to $10^{-4}\,\mathrm{m}^{-2}$ shrinks the feasible radius by the expected factor $\sqrt{10}$.

\subsection{Feasible density for 60 kbps IoT support}
\label{ssec:discussion_iot_density}

The second use case reverses the question: keep the $20\,\mathrm{km}$ footprint and ask what user density can be supported if each device
requires only $60\,\mathrm{kbps}$. Since $60\,\mathrm{kbps}$ is $0.06$ of $1\,\mathrm{Mbps}$, a user that would require $d$ RBs at
$1\,\mathrm{Mbps}$ would require about $0.06d$ RBs at $60\,\mathrm{kbps}$ under the same channel conditions. RB demand is integer-valued,
however, and a served user cannot consume less than one RB. We therefore map each raw $1\,\mathrm{Mbps}$ demand class $d$ to
\begin{equation}
\label{eq:discussion_iot_class_mapping}
d_{60}
=
\min\{20,\lceil 0.06d\rceil\},
\end{equation}
where the ceiling enforces an integer number of RBs and the minimum with $20$ keeps the same operator cap $M=20$.
This is a conservative class-level conversion rather than an exact recomputation of \eqref{eq:nrb_demand} at
$60\,\mathrm{kbps}$. The reason is that $d$ is already an integer RB class obtained after rounding up the $1\,\mathrm{Mbps}$ demand.
If we instead recomputed the continuous Shannon expression at $60\,\mathrm{kbps}$ and only then rounded to an integer number of RBs,
the resulting class could only be the same or smaller. Thus the conversion in \eqref{eq:discussion_iot_class_mapping} does not
understate the IoT budget. It gives a slightly conservative scaling estimate, which is sufficient for the feasibility comparison here.

Applying this class conversion to the short-range scenarios gives a new baseline budget
$B_{\mathrm{G}(\theta)}^{60\,\mathrm{kbps}}$ at $20\,\mathrm{km}$ and $\lambda_0=10^{-5}\,\mathrm{m}^{-2}$. If the density is changed from $\lambda_0$ to
$\lambda$, the footprint radius and per-device rate are now held fixed, so the budget scales approximately as
\begin{equation}
\label{eq:discussion_iot_density_scaling}
B_{\mathrm{G}(\theta)}^{60\,\mathrm{kbps}}(\lambda)
\approx
B_{\mathrm{G}(\theta)}^{60\,\mathrm{kbps}}
\frac{\lambda}{\lambda_0}.
\end{equation}
Setting this scaled budget equal to the satellite pool $B_{\mathrm{sat}}$ and solving for $\lambda$ gives
\begin{equation}
\label{eq:discussion_iot_feasible_density}
\lambda_{\max}(\mathrm{G}(\theta))
=
\lambda_0
\frac{B_{\mathrm{sat}}}{B_{\mathrm{G}(\theta)}^{60\,\mathrm{kbps}}}.
\end{equation}
Table~\ref{tab:discussion_iot_density} reports the resulting density estimates. The first numeric column is the
$60\,\mathrm{kbps}$ RB requirement at the baseline density $10^{-5}\,\mathrm{m}^{-2}$; the second numeric column is the maximum density
supported over the same $20\,\mathrm{km}$ footprint.
\begin{table}[!t]
\centering
\caption{Approximate feasible user density \(\lambda_{\max}\) for \(60\,\mathrm{kbps}\) IoT support over a fixed \(20\,\mathrm{km}\) footprint under the short-range attenuation scenarios.}
\label{tab:discussion_iot_density}
\small
\setlength{\tabcolsep}{4pt}
\begin{tabular}{lrr}
\hline
Attenuation & \(B^{60\,\mathrm{kbps}}\) & \(\lambda_{\max}\) \\
\hline
Stronger & \(23{,}343\) & \(7.4\times 10^{-6}\,\mathrm{m}^{-2}\) \\
Medium   & \(13{,}453\) & \(1.28\times 10^{-5}\,\mathrm{m}^{-2}\) \\
Mild     & \(13{,}066\) & \(1.32\times 10^{-5}\,\mathrm{m}^{-2}\) \\
\hline
\end{tabular}
\end{table}
The IoT case is therefore much less dominated by link rate and much more dominated by the number of devices. In the medium and mild
attenuation settings, the feasible density is only slightly above $10^{-5}\,\mathrm{m}^{-2}$ because a $20\,\mathrm{km}$ footprint at
that density already contains about $12{,}600$ expected users, and each of those users consumes at least one RB. In the stronger
attenuation setting, some devices require more than one RB even at $60\,\mathrm{kbps}$, so the feasible density falls below
$10^{-5}\,\mathrm{m}^{-2}$.

\section{Conclusion}
\label{sec:conclusion}

This work provides a pre-deployment dimensioning rule for satellite footprint RB capacity under spatially correlated attenuation. The rule
connects a footprint, a PPP user model, a link-budget-to-RB map, an operator per-user cap, and a Gaussian spatial attenuation family
to a footprint-level RB budget. The Monte Carlo procedure samples the Gaussian field only through its finite vector of values at the realized user
locations in each trial, which is the representation used to compute aggregate capped demand.

The empirical study shows that spatial covariance cannot be treated as a cosmetic propagation detail. With fixed marginal variance,
changing the correlation length from short range to footprint scale substantially changes the required RB budget and the variability of
total footprint demand. The strongest conclusion is therefore qualitative as well as numerical: a one-point attenuation law can describe
the marginal behavior of a single link while still missing the aggregate-load variance that determines the high-reliability footprint
budget.

The averaged-bound concentration calculation gives a secondary conservative check on the Monte Carlo budgets. It is not the main truth
model; it is an analytic upper bound for the same overload event. In the short-range cases, it closely tracks the simulation budget. In the footprint-scale
cases, it highlights the danger of relying on a single sampled field representation, because one realization can sit far from the
scenario-averaged behavior. If an operator is confident in the footprint, user density, attenuation family, link-budget assumptions, and cap,
the pipeline gives a direct high-reliability RB budget for that scenario; if those inputs are uncertain, the same pipeline gives a
controlled way to compare scenarios before deployment.

\section*{Acknowledgments}

OpenAI ChatGPT-5.4 was used to assist with code generation, the generation of Figs.~1--3, and grammatical editing of the
manuscript. The authors reviewed and validated the resulting code, figures, and text edits.

Laurent Decreusefond acknowledges support from the French National Research Agency (ANR) through project no.~ANR-22-PEFT-0010 of the France 2030
program PEPR Réseaux du Futur.

Philippe Martins acknowledges support from Bpifrance through the 5G NTN mmWave project.


\end{document}